\documentclass{ws-procs9x6}

\usepackage{graphics,epsfig}

\newcommand{\be}{\begin{equation}}
\newcommand{\ee}{\end{equation}}

\newcommand{\ba}{\begin{eqnarray}}
\newcommand{\ea}{\end{eqnarray}}

\begin{document}

\title{Pentaquarks: Theory overview, and some more about quark models}

\author{CARL~E. CARLSON\footnote{\uppercase{W}ork partially
supported by the \uppercase{N}ational \uppercase{S}cience \uppercase{F}oundation (\uppercase{USA}) under grant \uppercase{PHY}-0245056.}}

\address{Particle Theory Group, Physics Department\\
College of William and Mary\\ 
Williamsburg, VA 23187-8975, USA\\ 
E-mail: carlson@physics.wm.edu}

\maketitle

\abstracts{After reviewing the basics, topics in this talk include an attempted survey of theoretical contributions to this workshop, some extra specific comments on quark models, and a summary.}

%%%%%%%%%%%%%%%%%%%%%%%%%%%%%%%%%%%%%%%%%%%%%%%%%%%%%%%%%%%%%%%%%%%%%%%%

\section{ {Basics}}

%%%%%%%%%%%%%%%%%%%%%%%%%%%%%%%%%%%%%%%%%%%%%%%%%%%%%%%%%%%%%%%%%%%%%%%%

At the PANIC conference in Osaka in October 2002, a collaboration working at SPring-8 announced a baryon state\cite{Nakano:2003qx}, the $\Theta^+(1540)$ with a decay mode $\Theta^+ \to n K^+$.
Assuming the decay is not Weak, that meant it had strangeness +1, and therefore was an exotic state (``exotic'' in the sense that it is a baryon that cannot be a $q^3$ state).  It can be a $udud\bar s$ state: a pentaquark.  It has since been found at other labs by other collaborations, and also has failed to be seen in some experiments: an experimental review follows this talk\cite{hicks}.

In terms of flavor, considering $u$, $d$, and $s$ quarks, a quark is in a $\mathbf 3$ representation and an antiquark is in a $\bar{\mathbf 3}$, and the relevant product is
\ba
3 \otimes 3 \otimes 3 \otimes 3 \otimes \bar 3 
	= 1 \oplus 8 \oplus 10 \oplus \overline{10} \oplus 27 \oplus 35
\ea
(with multiplicities omitted on the right hand side).

\begin{figure}[h]

\hfil \includegraphics[width=1.5in]{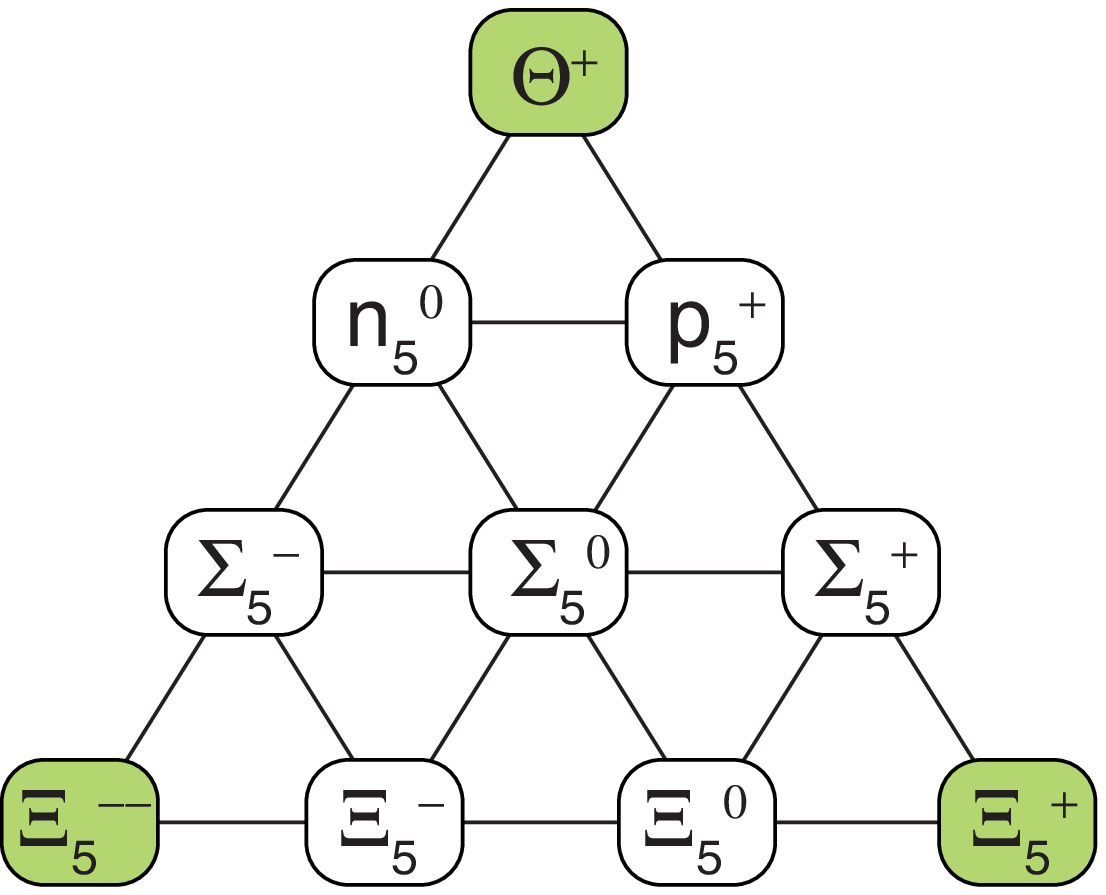}  \hfil
\includegraphics[width=1.5in]{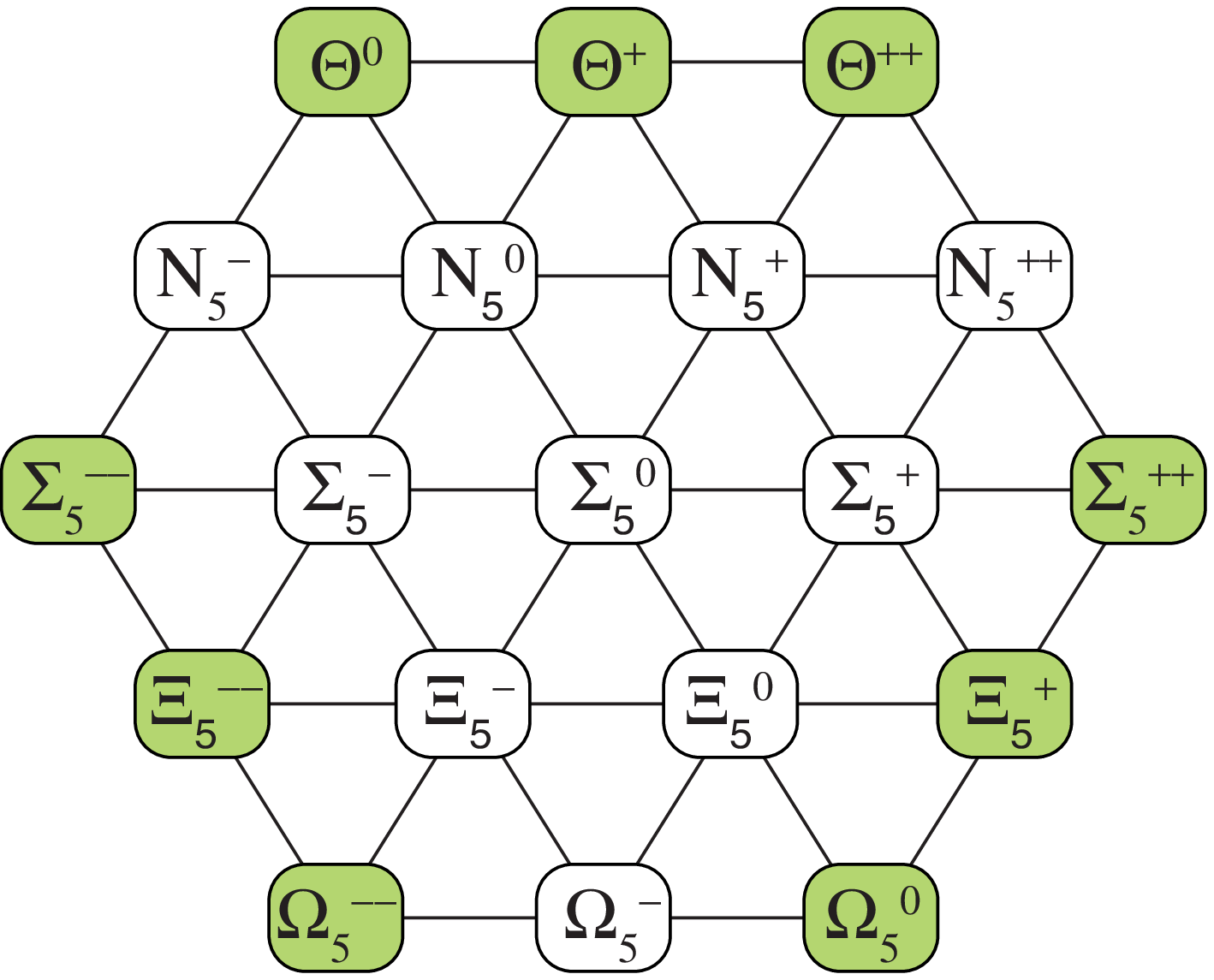}  \hfil

\caption{Two flavor multiplets with strangeness +1 states, the $\bar 10$ on the left and the $27$ on the right.  The shaded states are guaranteed to be exotic.  (Note: the Particle Data Group has proposed that the $I=3/2$, $S=-2$ be called $\Phi$ rather than $\Xi$.)}

\label{fig:exotic}

\end{figure}

An exotic must be in a $\overline{10}$, a $27$, or a $35$.  We have room to show two of these in Fig.~\ref{fig:exotic}; the $35$, not shown, has an isotensor multiplet (quintet) of $\Theta$'s.   Since searches for a $\Theta^{++}$ don't find it, we believe the $\Theta^+(1540)$ is isosinglet, and in a flavor $\overline{10}$.  The spin is not known experimentally; we  will assume it is 1/2.  Neither is the parity known, and we shall consider both possibilities.

%%%%%%%%%%%%%%%%%%%%%%%%%%%%%%%%%%%%%%%%%%%%%%%%%

%%%%%%%%%%%%%%%%%%%%%%%%%%%%%%%%%%%%%%%%%%%%%%%%%

%%%%%%%%%%%%%%%%%%%%%%%%%%%%%%%%%%%%%%%%%%%%%%%%%%%%%%%%%%%%%%%%%%%%

\section{Talks at this conference}

%%%%%%%%%%%%%%%%%%%%%%%%%%%%%%%%%%%%%%%%%%%%%%%%%%%%%%%%%%%%%%%%%%%%

I think it worthwhile to catalog the talks we heard at this workshop.  It gives some idea of the effort that has gone into understanding the pentaquark, and of the variety of starting points for that understanding.  (In some places I have also listed work not directly reported here.)

Let me start with lattice  gauge theory and with QCD sum rules, with a short comment following.

$\bullet$ Lattice gauge theory.  There were talks at this conference by Chiu, Ishii, Takahashi, Sasaki, and F. Lee.   The table notes what parity the pentaquark was found to have by each collaboration---if they saw a pentaquark signal at all.

\begin{tabular}{|ll|}
\hline
\hline
\quad Mather (Kentucky) {\it et al.} & \quad no pentaquark \quad  \\
\quad Chiu and Hsieh (Taiwan)  &  \quad + parity  \\
\quad Sasaki (Tokyo) &  \quad -- parity   \\
\quad Csikor {\it et al.} (Hungary/Germany) \quad  & \quad  -- parity \\
\hline
\quad MIT \quad  & \quad  -- parity \\
\quad Ishii {\it et al.}  \quad  & \quad  no pentaquark \\
\quad Takahashi {\it et al} \quad  & \quad  -- parity \\
\hline
\end{tabular}                      \bigskip

%%%%%%%%%%%%%%%%%%%%%%%%%%%%%%%%%%%%

$\bullet$  QCD sum rules  \bigskip

\begin{tabular}{|ll|}
\hline
\quad Sugiyama, Doi, Oka \quad & \quad negative parity \quad \\
\quad Nishikawa  \quad & \quad positive parity  \\
\quad S. H. Lee  \quad &  \quad negative parity again \quad  \\
\hline
\end{tabular}   \bigskip

Lattice gauge theory and QCD sum rules share a problem (of which the practitioners are well aware):  The $\Theta^+$ state we want lies within an $NK$ continuum.  The threshold for the continuum is {\it below} the $\Theta^+$.  That means the desired signal is exponentially suppressed relative to continuum.  Part of the practitioners skill is in choosing test operators with little overlap with the continuum.  Hence the discussions---and also hope of progress.

%%%%%%%%%%%%%%%%%%%%%%%%%%%%%%%%%%%%%

$\bullet$ Chiral soliton or Skyrme models.  These were the original motivational springboard of the modern pentaquark era.  We should mention the work of Diakonov, Petrov, Polyakov; of Manohar;  of Weigel; of Praszalowicz;  of Hirada;  of Chemtob;  of Oh, Park, \& Min;  a good fraction of these workers were represented here.

%%%%%%%%%%%%%%%%%%%%%%%%%%%%%%%%%%%%%%%

$\bullet$  Quark model.  The quark model allows one to explicit calculations of many pentaquark properties.  There were contributions regarding wave functions, masses, production rates, magnetic moments, etc., and I also list a few authors who were not represented here.

{\normalsize 
\begin{tabular}{|ll|}
\hline
Quark cluster states & H{\o}gaasen and Sorba \\
	& Lipkin and Karliner \\
	& Jaffe and Wilczek \\
Consequences of Flavor-Spin \qquad \qquad & Riska and Stancu \\
 hyperfine interactions	& Jennings and Maltman;  \\
	& C., Carone, Kwee, Nazaryan \\
Real 5q calculations  & Takeuchi;  Enyo; Hiyama \\
\ \ \ (given a choice of $\mathcal H$) 
	&	Maezawa (w/movies!);   Okiharu  \\
Instanton motivated potentials & Shinozaki  \\
Higher representations	& Dmitrasinovic  \\
	& Y. Oh \\
	& Manohar \& Jenkins \\
Magnetic moments &  H. C. Kim  \\
Production  &  S. I. Nam;  \qquad C. M. Ko  \\
\hline
\end{tabular}
}

%%%%%%%%%%%%%%%%%%%%%%%%%%%%%%%%%%%%%%%%%

$\bullet$ Consequences of strong coupling of $\Theta$ to nucleon + {\sl two} mesons.  Under this heading comes the possibility that the $\Theta$ itself is a ``molecular'' state, {\it i.e.}, an $NK\pi$ w/40 MeV binding, as well as binding of the $\Theta$ within a nuclear medium.  Workers here include Vacas, Nagahiro, Bicudo, Oset, Kishimoto, and Sato.

And to close this catalog, we have

$\bullet$  String views of pentaquarks, reported by Suganuma and by Sugamoto.

$\bullet$ Methods of parity measurement---theory. Discussed here by Hanhardt.

$\bullet$ Why pentaquarks are unseen at high energy.  Discussed by Titov.

We continue with some additional remarks in the context of quark models of pentaquarks.  An important question is the parity of the state.

%%%%%%%%%%%%%%%%%%%%%%%%%%%%%%%%%%%%%%%%%%%%%%%%%%%%%%%%%%%%%%%%%%%%%%%%

\section{Negative parity}

%%%%%%%%%%%%%%%%%%%%%%%%%%%%%%%%%%%%%%%%%%%%%%%%%%%%%%%%%%%%%%%%%%%%%%%%

\begin{figure}
\label{fig:q5}

\centerline{  \includegraphics[height=2.cm]{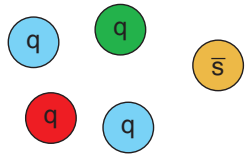}  }

\caption{Pentaquark state.}
\end{figure}

The easiest $q^4 \bar q$, Fig.~\ref{fig:q5}, state to consider in a quark model is one where all the quarks are in lowest spatial state.  Since the $\bar q$ has negative intrinsic parity, the state overall will have negative parity.  This can be good in that it may be supported by lattice calculations and QCD sum rules, or bad in that it disagrees with results of chiral soliton models, or really bad in that a straightforward version will lead to very broad decay width.

It is, of course, crucial that the $q^4$ part of the pentaquark wave function be totally antisymmetric, as required by Fermi statistics.  It is useful to write out the full state.  One may choose different ways to express it, and we will display the result of breaking the state into a sum of terms where each term is a product of a $q^3$ state and a $q \bar q$ state.  For the $q^4 \bar q$ state that has the quantum numbers of the $\Theta^+$, the result is\cite{Carlson:2003pn}
\ba 
| \Theta^+ \rangle =
\frac{1}{2}  | NK  \rangle
	+ \frac{1}{\sqrt{12}}  | NK^* \rangle
		+ \sqrt{\frac{2}{3}} 
	\ \Big| ({\rm color\ octet}) * ({\rm color\ octet}) \Big\rangle   .							
\ea

\noindent   The coefficients are fixed uniquely (up to a phase) by the requirement of antisymmetry when a quark in the $q^3$ is exchanged with the quark in the $q\bar q$.

The notation ``$N$'' in the above expansion means a state with the color, flavor, and spin structure of a nucleon.  The spatial wave function may be different, for example, it may be more spread out than for an isolated nucleon state: it is, after all, embedded in a $\Theta^+$.  Similar comments hold for the $K$ and $K^*$.  However, the $NK$ quantum number part of the state is the one that, at least naively, can easily fall apart into a real nucleon plus kaon final state.  The other pieces of the state, when taken apart, either weigh more than the $\Theta^+$ or are colored.  Hence only 25\% of state can decay into kinematically allowed final states (12$\frac{1}{2}$\% each for $n K^+$ and $p K^0$).  We shall see that 25\% is a big number, after we discuss the positive parity case.

%%%%%%%%%%%%%%%%%%%%%%%%%%%%%%%%%%%%%%%%%%%%%%%%%%%%%%%%%%%%%%%%%%%%%%%%

\section{Positive parity}

%%%%%%%%%%%%%%%%%%%%%%%%%%%%%%%%%%%%%%%%%%%%%%%%%%%%%%%%%%%%%%%%%%%%%%%%

Positive parity requires that one of the quarks be in a P-state excitation.
One then expects the state to get heavier and one has to discuss counter mechanisms that could keep the state light.

\begin{figure}
\label{fig:cluster}
\centerline{  \includegraphics[height=3cm]{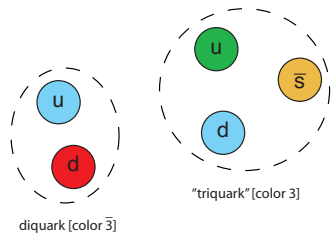}  }
\caption{Clustered quark possibility}
\end{figure}

One countermechanism is to cluster the quarks, and ignore interactions among quarks in different clusters.  Quarks within a given cluster do have color-spin interactions.  An illustration with two clusters is shown in Fig.~\ref{fig:cluster}.  

%%%%%%%%%%%%%%%%%%%%%%

%%%%%%%%%%%%%%%%%%%%%%%%%%%%%%

One seeks cluster possibilities that have maximal binding.  Ignoring intercluster interactions, it can be made to work.  For example, looking at the figure, choose the diquark as  C = $\bar 3$, F = $\bar 3$, S = 0, and the $qq\bar q$ ``triquark'' as  C = 3, F = $\bar 6$, S = 1/2, and put the clusters in a relative spatial P-wave.  This is a suggestion of Karliner and Lipkin; one should see\cite{Hogaasen:2004ij} and a version by Jaffe and Wilczek with clustering into diquark-diquark-$\bar q$\cite{Hogaasen:2004ij}.

However the requirement to keep the clusters separated so that intercluster $qq$ and $q \bar q$ interactions small---and to dare neglect Fermi statistics requirements of quarks in separate clusters---seems dangerous.

We wish also to consider uncorrelated states, with Fermi statistics fully satisfied.  One the needs different mechanism to make positive parity lighter than negative parity, since the color-spin attractions among quarks prove unsuccessful.  However, another possibility is at hand, in the form of flavor-spin interactions for quark pairs.  Consider the mass shift operator,
\ba
\Delta M_\chi = -C_\chi \sum_{\alpha<\beta} 
	\left( \lambda_F \sigma \right)_\alpha  \cdot
	\left( \lambda_F \sigma \right)_\beta  \ .
\ea

This flavor-dependent mass operator was originally put forward because it had good effects on $q^3$ baryon spectroscopy\cite{Glozman:1995fu}.  Later comparative studies of mass operators for excited states indicated flavor-spin operators were necessary to describe the observed mass spectrum\cite{Carlson:1998gw}.

%  There is also the Manohar-Jenkins result that, in the large $N_c$ limit, %states in the chiral soliton model can be mapped to fully flavor-spin symmetric %quark model states, which are precisely the states favored by the above mass %operator.

Be all that as it may, the question remains:  For the $q^4$ part of state, how can an $S^3 P$ configuration be lighter than an $S^4$?

\begin{figure}
\label{fig:q3}
\centerline{  \includegraphics[height=1.9in]{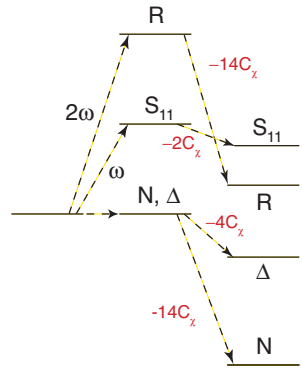}  }
\caption{Some $q^3$ baryons.}
\end{figure}

Recalling the $q^3$ baryon case is useful.  It shows the mass advantage of a totally symmetric flavor-spin state (the relevance of this to the pentaquark will become clear) and sets the value of the parameter $C_\chi$.  The problem in the $q^3$ sector is that the Roper or $N^*$(1440) [an excited S-state] is lighter than the P-states, {\it e.g.}, the $S_{11}$(1535).   With spin-dependent interactions, the situation is like the intermediate part of Fig.~\ref{fig:q3}, drawn for a potential like the harmonic oscillator.  The Roper is heavier than the nucleon by two oscillator spacings, the P-state by only one.  However, the Nucleon,  Roper, $\Delta$ all have symmetric flavor-spin wave functions, and the spin-dependent interactions can pull them down much more that the $S_{11}$.  Explicitly,
\be
\Delta M_\chi = \left\{ 
	\begin{array}{cl}
	-14 C_\chi & \quad N(939), N^*(1440) \\
	 -4 C_\chi & \quad \Delta(1232) \\
	 -2 C_\chi & \quad N^*(1535)
\end{array} \right.   \,
\ee

\noindent and the masses become like the final part of Fig~\ref{fig:q3}.  One picks $C_\chi = 30$ MeV from $N$-$\Delta$ mass splitting, and chooses an oscillator spacing $\hbar\omega \approx 250$ MeV.

For the pentaquark, the $S^3P$ state has, or can have, a totally symmetric flavor-spin wave function, the $S^4$ cannot.  The effect is large:
\ba
M(S^3 P) - M(S^4) = \hbar\omega - \frac{56}{3} C_\chi 
	\approx -310 {\rm\ MeV}  \ . 
\ea

\noindent The flavor-spin mass interaction makes positive parity pentaquark states lighter than the negative parity ones.

As for the negative parity case, we can write the $q^4 \bar q$ positive parity  state as a unique sum of products of $q^3$ and $q \bar q$ states, where the $q^3$ or $q \bar q$ may have the quantum numbers of observed hadrons, or may (more often) be color octet states.  The details are in\cite{Carlson:2003xb,Carlson:2003wc}.  The only $q^3$ and $q \bar q$ combination with quantum numbers allowing a color and kinematically possible decay is the $NK$ combination, and the overlap is small.  We quote
\be
		| \ \langle NK | \Theta^+ \rangle \ |^2 = \frac{5}{96} \ .
\ee

%%%%%%%%%%%%%%%%%%%%%%%%%%%%%%%%%%%%%%%%%%%%%%%%%%%%%%%%%%%%%%%%%%%%%%%%

\section{ {Narrowness}}

%%%%%%%%%%%%%%%%%%%%%%%%%%%%%%%%%%%%%%%%%%%%%%%%%%%%%%%%%%%%%%%%%%%%%%%%

Given the overlap just calculated, one could simply estimate that for the positive parity case, $\Gamma(\Theta^+)$ is 5\% of typical strong interaction decay.

One can do better by invoking effective field theory ideas, using a Lagrangian (for the positive parity case)
\ba
{\mathcal {L}}_{int} = ig_+ \ \bar \psi_N \gamma_5 \psi_P \ \phi  
+ {\rm hermitian\ conjugate}.
\ea
One gets the naive size of the coupling parameter from naive dimensional analysis, and modifies it with using the just calculated $\Theta$\,--$NK$ overlap, as 
$g_+^2 = P_{overlap} \big( g_+^{(naive)} \big)^2$.  The consequence of naive dimensional analysis is $g_+^{(naive)} = 4\pi$, incidentally the same---quite accurate---result one obtains for $g_{\pi NN}$.

Working out the phase space for the positive parity $\Theta^+$, one gets\cite{Carlson:2003xb,Hosaka:2004bn}
\be
\Gamma_+ = \frac{5}{96} \times 85 {\rm\ MeV} = 4.4 {\rm\  MeV}  \,.
\ee

\noindent  The overlaps taken into account are only for the color-flavor-spin part of the wave function.  The physical size of the spatial wave functions of the $\Theta^+$, nucleon, and $K$ are surely not the same, and taking overlaps of spatial wave functions also into account will {\it reduce} the above estimate, possibly significantly.

For the negative parity state, the bigger $P_{overlap}$ and bigger phase space (S-wave rather than P-wave) leads to the astonishing result
\be
\Gamma_- \approx 1.0 {\rm\ GeV}.
\ee

%%%%%%%%%%%%%%%%%%%%%%%%%%%%%%%%%%%%%%%%%%%%%%%%%%%%%%%%%%%%%%%%%%%%%%%%

\section{ {Closing comments}}

%%%%%%%%%%%%%%%%%%%%%%%%%%%%%%%%%%%%%%%%%%%%%%%%%%%%%%%%%%%%%%%%%%%%%%%%

There has been lots of theoretical work on pentaquarks, from assorted starting points.  It was exciting to hear the various talks presented here.

Lattice gauge theory and QCD sum rule workers reported a number of independent calculations.   The majority found a signal for the pentaquark, and found it with negative parity.  But right now it would be safer to say that the parity and existence of pentaquarks not settled within the context of these two techniques. 

Chiral quark soliton models (still) predict the existence of pentaquarks, and put the lightest one in a flavor $\overline {10}$ with  mass in 1500 Mev range and with positive parity.

The quark model can accommodate either parity.  It seems easier to explain the narrowness of the state if the parity is positive.  Numerical estimates of the widths of positive and negative parity pentaquark are dramatically different, and well justified estimates of the positive parity widths are as narrow as the experiments are requiring.

Then there is the possibility that $\Theta^+$ is $NK\pi$ bound molecular bound state.  Let me describe this idea as intriguing, and note that it could render much of the other work moot.

There are still things to do.  In particular, one thing somewhat but not heavily represented here were studies of excited pentaquarks.  \medskip

Greatest thanks to all the organizers of this very excellent workshop.

\end{document}